\documentclass[runningheads]{llncs}
\usepackage{graphicx}
\usepackage{paralist}
\usepackage{comment}
\usepackage{pdflscape}
\usepackage{nameref}
\usepackage{multicol}
\makeatletter
\@mparswitchfalse
\makeatother

\usepackage[textsize=scriptsize,textwidth=4cm,disable]{todonotes}

\usepackage{xcolor}
\usepackage[spaces,hyphens]{url} 
\usepackage[colorlinks=false,urlcolor=blue]{hyperref} 

\begin{document}
%
\title{A Norm Emergence Framework for Normative MAS~-- Position Paper}
%
%
\author{Andreasa Morris-Martin\inst{1}
\and
Marina De Vos\inst{1} 
\and\\
Julian Padget\inst{1} \orcidID{0000-0003-1314-2094}
}

%
\authorrunning{A. Morris-Martin et al.}
%
\institute{University of Bath, Bath, UK
\email{\{a.l.morris.martin,m.d.vos,j.a.padget\}@bath.ac.uk}}
\maketitle              
\begin{abstract}
Norm emergence is typically studied in the context of 
multiagent systems (MAS) where norms are implicit, and participating agents use simplistic decision-making mechanisms. These implicit norms are usually unconsciously shared and adopted through  
agent interaction. A norm is deemed to have emerged when a threshold or predetermined percentage of agents follow the ``norm''. 
Conversely, in normative MAS, norms are typically explicit and agents deliberately share norms through communication or are informed about norms by an authority, 
following which an agent decides whether to adopt the norm or not.  
The decision to adopt a norm by the agent can happen immediately after recognition or when an applicable situation arises.
In this paper, we make the case that, similarly, a norm has emerged in a normative MAS when a percentage of agents adopt the norm. 
Furthermore, we posit that agents themselves can and should be involved in norm synthesis, and hence influence the norms governing the MAS, in line with Ostrom's eight principles. 
Consequently, we put forward a framework for the emergence of norms within a normative MAS, that allows participating agents to propose/request changes to the normative system, while special-purpose synthesizer agents formulate new norms or revisions in response to these requests. 
Synthesizers must collectively agree that the new norm or norm revision should proceed, and then finally be approved by an ``Oracle''. The normative system is then modified to incorporate the norm.

\end{abstract}

\section{Introduction}
Multiagent systems (MAS) enable participating agents to interact with each other and their environment to accomplish individual goals and collective goals. MAS utilise norms to encourage coordination and cooperation, and to avoid the occurrence of undesirable states. 
Hollander and Wu~\cite{DBLP:journals/jasss/HollanderW11} define a normative MAS as a system which combines concepts of norms with explicit representations of normative information in order to provide a solution to problems relating to openness in MAS. In contrast, in some MAS norms may not be considered at all or the concept of norms is present but only as implicit representations of normative information.

The normative system, the set of norms in the MAS, is normally created as a result of the process of norm synthesis. Norm synthesis is the creation and updating of norms to avoid conflict situations~-- unwanted states in the MAS. 
Norm synthesis can be offline, which occurs mostly during design time \cite{DBLP:conf/dalt/LeePLDA14} or as a separate process outside the system governed by the norms \cite{morales_off-line_2018}, where norms are determined by the designer or other stakeholders. 
A normative system resulting from offline synthesis is typically fixed for the lifetime of the system. But, over time, with changing environments, the norms can become partly or wholly irrelevant. Consequently, it becomes necessary to update the normative system leading to the introduction of online norm synthesis. 
This occurs while the system is live, and (new/revised) 
norms are typically determined by a centralised mechanism with global knowledge and without any input from the participants~\cite{Mashayekhi:2016:SSS:3060621.3060674,Alechina:2015:PRN:2772879.2772937}. 
%
Conversely, the introduction of norms into a MAS, without an explicit normative system, 
is orchestrated by participating agents in simulation models of norm emergence~\cite{villatoro_topology_2009,mukherjee2007emergence,savarimuthu_norm_2009}, where the norms are defined as the preferred action from a set of actions, and the norm is usually learnt through agent interactions. 

In this paper, we introduce a framework that allows participating agents to affect the online norm synthesis process. 
During runtime, participating agents can identify the norms or situations that require normative regulation, as long as there are the necessary affordances \cite{DBLP:conf/atal/NoriegaSVPd17}  
for participating agents to propose or request changes to the normative system.  
We propose synthesizer agents which, upon request from participating agents, synthesise norms for the MAS to meet the identified need. 
We believe that after we encode the synthesised norm into the normative system, and it is adopted by a sufficient number of agents, we may ultimately observe the emergence of synthesised norm(s). 

The rest of the paper is structured as follows.  
Section~\ref{context} which situates the proposed framework by discussing the gap identified in the literature. Section~\ref{description} gives a high level description of the framework, after which we discuss the stages of the framework in more detail: Section~\ref{normCreate} highlights the main contribution which defines the norm creation stage, Section~\ref{normProp} discusses norm propagation, then norm adoption in Section~\ref{normAdopt}, finishing in Section~\ref{normEmer}, with norm emergence.  
The paper concludes in Section~\ref{conclude}, where we briefly discuss 
how the framework can facilitate the emergence of norms in open normative MAS and allow for the normative component to be changed in response to issues affecting agents participating in the MAS.




\section{Research Context}
\label{context}
Norms in the (MAS) literature are looked at from two distinct perspectives: norms as deontic concepts and norms as a preference behaviour. Each perspective implements the life cycle of norms differently, but can broadly be seen to follow similar stages. The norm life cycle defines the stages a norm goes through in its lifetime. 
Hollander and Wu~\cite{DBLP:journals/jasss/HollanderW11} define a norm life cycle consisting of: creation, transmission, recognition, enforcement, acceptance, modification, internalisation, emergence, forgetting and evolution. These processes are encapsulated into three super processes: enforcement, internalisation and emergence with evolution defined as an end-to-end processes.  
The widely accepted norm life cycle \cite{savarimuthu_norm_2011}, based on simulation studies has a three stage process: norm formation/creation, norm propagation and  norm emergence. 
A closer look at the norm life cycle enables us to better understand the different perspectives on norms. 

\subsection{The prescriptive perspective and the norm life cycle}
The first perspective, norms as deontic concepts, views norms as permissions or prohibitions and obligations, or commitments and is referred to as \emph{norms as prescriptions\/}~\cite{conte_conventions_1999}, or the \emph{prescriptive approach\/}~\cite{DBLP:conf/iat/SavarimuthuAS11}. Here we refer to it as the \emph{prescriptive perspective}.
Its literature uses an explicit representation of norms, typically internal as agent beliefs, but can also be external and referenceable. 

The norm life cycle in a society with explicit norm representation exhibits the following stages:
\begin{inparaenum}[(i)]
\item creation: a norm is introduced into the society, source often unknown,
\item identification: agents become aware of the norm and update their beliefs,
\item adoption: agents reason about adopting the norm,
\item propagation: agents deliberately, or possibly unintentionally, inform others of the norm,
\item emergence: not usually considered, but inferrable when a predetermined threshold of agents have adopted the norm.   
\end{inparaenum}

The prescriptive perspective branch of the literature presents the norm life cycle with a focus on norm adoption. 
It also considers the synthesis of norms as an alternative to the norm creation stage, as it defines the norms that will regulate the behaviour of the agents within the MAS. Norm synthesis is geared towards creating norms to avoid conflicts and can occur both online and offline. 
Recently in normative MAS, online norm synthesis is becoming a popular topic, as it allows for the introduction of norms into the MAS at runtime to meet changing circumstances.
Another benefit of online norm synthesis is that it can cater for observed states, or agent capabilities that had not been conceived at the time the MAS was designed~\cite{haynes_luck_mcburney_mahmoud_vitek_miles_2017}.
The online norm synthesis mechanisms cited \cite{morales2013automated,Morales:2015:SLN:2772879.2772936,morales_online_2015,Alechina:2015:PRN:2772879.2772937,Mashayekhi:2016:SSS:3060621.3060674,campos_robust_2013} 
here operate by monitoring the state of the environment and proposing norms to resolve conflicts, that can or have occurred, to prevent their occurrence in the future. For example, \cite{morales2013automated,Morales:2015:SLN:2772879.2772936,morales_online_2015} do so by prohibiting the action of an agent in the timestep before the conflict.  
The majority of online methods cited utilise a centralised mechanism with global knowledge, and to the best of our knowledge only AOCMAS~\cite{campos_robust_2013} employs a decentralised mechanism.

AOCMAS is a two-level 
distributed MAS architecture that equips an organisation with adaptive capabilities. At one level there are domain agents who are concerned with their individual goals, and above them are assistant agents who are concerned with the organisation's goals and help to facilitate the adaption. Assistants are responsible for or oversee a cluster of domain agents.
Assistant agents partially observe the state of the organisation at runtime and propose regulations, rules or legal norms, for problems identified. 
Before proposing new solutions, assistants check if an existing stored solution is applicable using case-based reasoning (CBR). If none is found, they propose a set of regulations, and each regulation in the set is voted on by all assistants. The regulations with the majority vote become the new set of regulations, that over time are evaluated for effectiveness and may be updated or removed.

We note that the literature on norm synthesis, which is applicable to the norm creation stage of the norm life cycle, fails to investigate the impact of the remainder of the norm life cycle. Instead the focus is on determining an appropriate set of norms, after which it is assumed that either these norms will be adopted by agents or that the norms will be regimented. 
It is surprising that a disconnect exists within a single perspective of the norm literature, but research in different sections of the literature investigate various processes of the norm life cycle independently, 
with little  attempt to combine or sequence the activities together.
This gap, we believe, can readily be filled, by using norm synthesis in lieu of norm creation, while the other stages continue as normal in the prescriptive approach. 

\subsection{The emergence perspective and the norm life cycle}
The second perspective of norms is as a preference behaviour. This views norms as a predetermined or computed preference behaviour to execute from among a set of behaviours in a given situation. 
The preference behaviour perspective is referred to as \emph{norms as conventions\/}~\cite{conte_conventions_1999} or the \emph{emergence approach\/}~\cite{DBLP:conf/iat/SavarimuthuAS11}, here we refer to it as the \emph{emergence perspective}. Its literature typically uses an implicit norm representation. 

The norm life cycle in a society with implicit norms usually has these stages:
\begin{inparaenum}[(i)]
\item creation: the initial or predetermined strategy for an agent from a set of available actions, 
\item propagation: the sharing of an agent's strategy either unintentionally or deliberately,
\item emergence: a percentage of agents follow this strategy for every instance of the triggering situation. \end{inparaenum}
The norm life cycle with implicit norms is usually seen in simulation models of norm emergence. 
Savarimuthu and Cranefield~\cite{savarimuthu_norm_2011} examine simulation models of norms in MAS and present an expanded five stage norm life cycle: creation, identification, spreading, enforcement and emergence. They also categorise various norm-based simulation works into each stage of the life cycle based on the mechanisms employed in each. 

The emergence perspective literature is focused primarily on the study of norm emergence. 
Norm emergence is normally defined as the point in a MAS when a threshold or predetermined percentage of agents adopt a norm. There is generally no discussion of the activities that precede or follow this point in the state of the MAS. In~\cite{morris-martin_norm_2019} we present the notion that a refinement of norm emergence is needed, as it is difficult to fully understand the emergence of a norm without taking into account the preceding activities of norm creation and propagation. 
Therefore norm emergence is henceforward what is normally the norm life cycle, but including norm creation and propagation. 

Utilising the interactions of agents at runtime as a basis for the creation of norms, as in norm emergence, is another potential solution for the source of norms in the prescriptive perspective. This is not an automatic fix however, as there are disadvantages to using norm emergence alone. 
Morales et al.~\cite{morales_off-line_2018} suggest that norm emergence is inappropriate to synthesise norms for MAS where there are numerous inter-dependent conflict situations. Additionally, norm emergence does not allow for the explicit representations of norms. 
This can be problematic, because when norms are implicit, there can be confusion among the agents about the prevailing norm, because different subsets of agents may adopt different norms. Additionally agents can have different interpretations about the prevailing norm(s) based on their beliefs \cite{DBLP:journals/jasss/HollanderW11}.  

\subsection{Research Opportunity}
As we have alluded to earlier, norm emergence in its traditional sense is comparable to online norm synthesis, since both involve the creation of norms at runtime. 
In norm synthesis the creation of the norms is motivated by a single external entity: external in the sense that it does not involve participating agents, while in norm emergence it is achieved internally by the participating agents' behaviour.
The creation of norms in norm emergence is naturally distributed and online, while norm synthesis is usually centralised and often times offline. 

We submit that it can be beneficial to use participating agents within the MAS in the norm synthesis process, thereby taking the benefits of the distributed and online approach of norm emergence and adopting it into the norm synthesis process, to produce explicitly represented norms. The final product is an approach that facilitates the emergence of norms in a normative MAS system. 
Building on the concept of distributed agents having the ability to create norms, we put forward a framework that allows participating agents to request the inclusion/modification of norms in the normative system, based on their experiences within the MAS.

We situate this notion of participating agents contributing to the norms on two of Ostrom's \cite{ostrom:1990} eight principles for governing the commons. Foremost, the need to allow participatory decision making, especially from those who are likely to be affected by the rules/decisions, and secondly the need to ensure the rules in place meet the needs of the local context, that is the participant's needs. We believe these two principles, which are intended to aid proper governance of the commons, are applicable to defining rules for normative MAS with different contexts. 

Haynes \emph{et al.}~\cite{haynes_luck_mcburney_mahmoud_vitek_miles_2017} examine how emergent behaviours in a system can be beneficial to the system and should be encouraged and spread, while non-beneficial emergent behaviour should be discouraged. We suggest that beneficial emergent behaviour can further be encoded explicitly as legal norms within the normative MAS. Therefore beneficial emergent behaviours can give rise to the emergence of obligation norms, permission norms and prohibition norms for alternative 
behaviours. Likewise, non-beneficial emergent behaviours could give rise to the emergence of prohibition norms, obligation norms to avoid certain states and revocation of permission norms.

Note that we do not expect participating agents to be able to properly synthesise norms for the MAS, because they would need access to domain knowledge and history of the MAS, 
which a participating agent would not normally have.   
Additionally, there is a need for higher minimum requirements on the cognitive abilities of participating agents that may make the MAS inaccessible to some agent types. 
Therefore, we propose agents communicate requests to synthesizer agents, who are responsible for a subset of agents within the MAS. Each synthesizer agent is capable of synthesising norms, based on the request from the participating agents, and their (partial) knowledge of the domain and the environment. 
We contend that the request for norm change from a participating agent aligns with triggering the initiation of norm synthesis from the bottom up.

\section{Norm emergence framework} 
\label{description}
\begin{figure}[t!]
\centering
\includegraphics[width=0.7\textwidth]{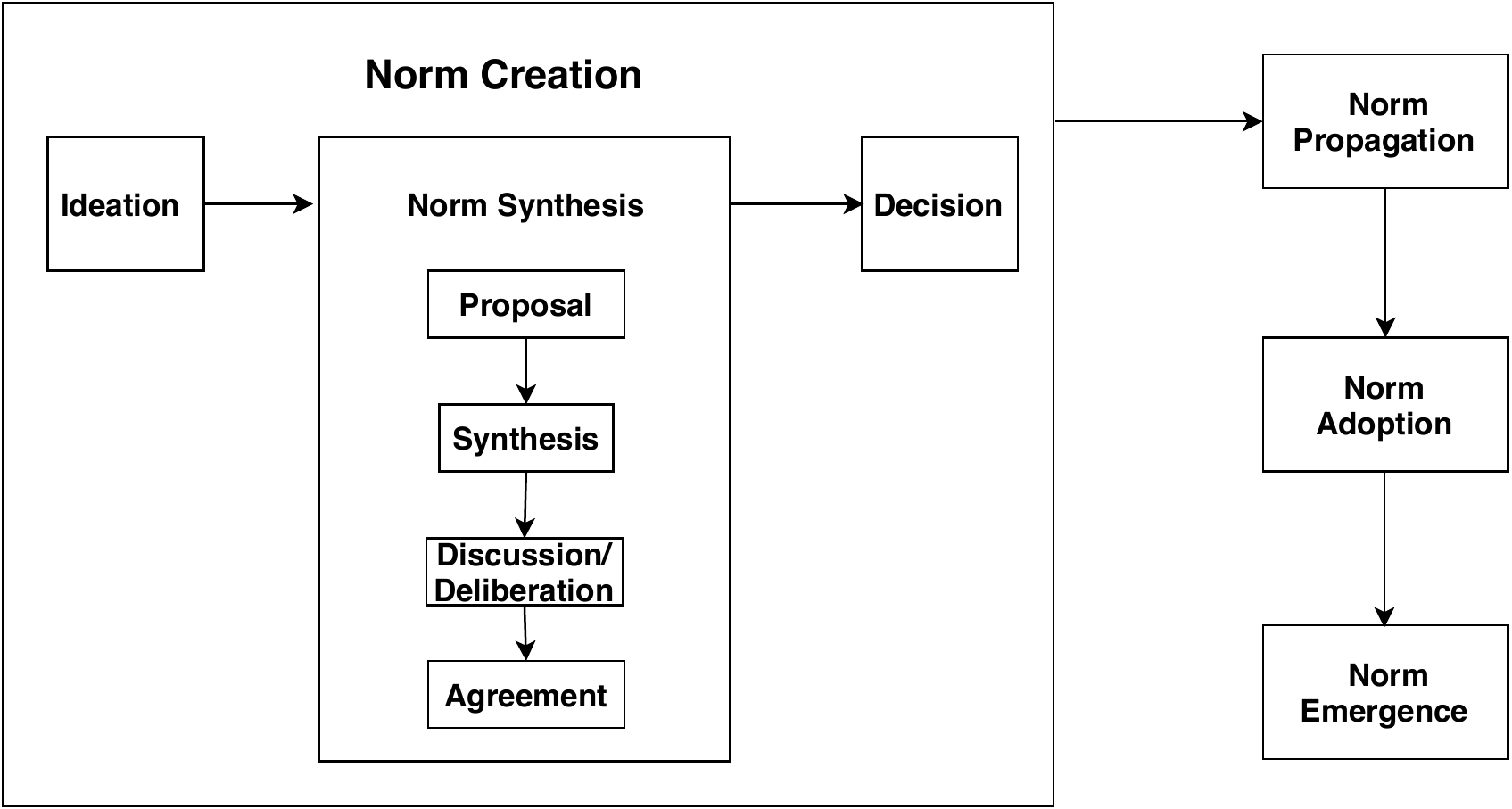}
\caption{A conceptualisation of norm emergence for normative MAS} \label{fig:summary}
\end{figure}

Agents operating in a MAS, like in human societies, could
potentially determine the norms that would better regulate the society. 
In the emergence perspective literature, it is accepted that agents learn their behaviour from interacting with other agents. This means that the strategy of one agent or a set of agents can become the norm in the society. The usefulness of employing the concept of the agents determining the norm, as a potential answer to the source of the norm in the prescriptive perspective, is worth investigating, thereby allowing the norms of a system to be determined by the agents that participate in the system.  

The goal of this approach is to explore techniques for developing self-governing systems, in which agents participate in the revision of the norms that affect them.
This would preclude the need for direct human intervention to synthesise norms for changing environments, or at the least, require minimal human intervention. The removal of human involvement though useful in this context unearths risks that must be considered in developing MAS. We provide a brief explanation in Section~\ref{decision} \nameref{decision}. 

\begin{figure}[p!]
    \advance\leftskip-2cm
    \rotatebox{90}{\begin{minipage}{\textheight}
    \includegraphics[keepaspectratio,width=\textwidth]{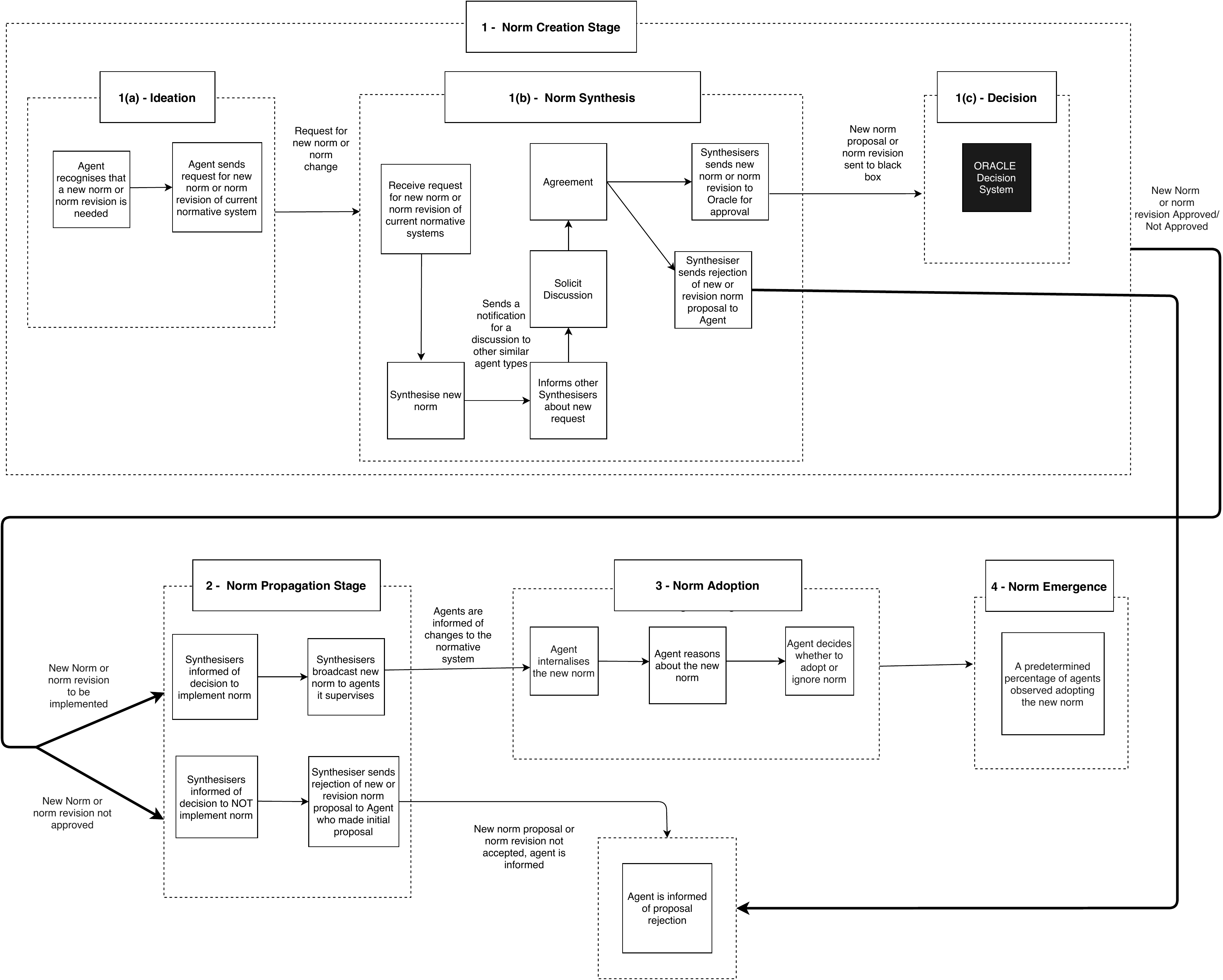}
     \caption{Conceptual Model for norm emergence in normative MAS}
     \label{fig:model}
     \end{minipage}}
  
  \end{figure}

We conceptualise a model that pulls together three research strands, namely: 
\begin{inparaenum}[(i)]
\item the life cycle of norms in the prescriptive perspective 
\item the life cycle of norms in the emergence perspective, and 
\item the norm synthesis process 
\end{inparaenum}, in order to construct a framework that facilitates the emergence of norms in a normative MAS. 
Thereby, we aim to identify the complementary activities~-- norm creation in emergence and norm identification and adoption in normative MAS and explicitly represented norms from norm synthesis, remove duplicate and unnecessary activities, and sequence them in a way that a new model is revealed. 
The resulting model enables us to define a norm emergence framework for normative MAS which we sketch in Figure~\ref{fig:model}. 

The model details the process of norms being synthesised, which begins with the initial recognition of the need for a new norm in the MAS by a participating agent. We conceive of two types of agents operating within the MAS: participant agents and synthesizer agents. The topology of the agents within the MAS will not be considered as we do not consider interactions within the MAS to be influenced by social distance or connections between participating agents. 
We will however aim for there to be a uniform distribution of agents to synthesizers based on the number of synthesizer agents in use.  

\medskip
\noindent \textbf{Participating Agents}
We put forward some assumptions about the agents that may participate in such a MAS, specifically: 
\begin{inparaenum}[(i)]
\item the agent has an explicit internal representation of norms and some non-trivial cognitive or reasoning abilities, 
\item the agent is capable of perceiving a need for a norm change, either as a new or revised norm, and
\item the agent considers norms in their action-selection and planning processes. 
\end{inparaenum} 

The agent's perception of the need for a norm change is predicated on agents considering norms when acting, and being able to observe the effect of their actions on the environment via action feedback. The participating agents in our model are inherently \textit{normative agents}. 

\medskip
\noindent \textbf{Synthesizer Agents}
The framework calls for a distinguished set of agents with partial perception of the MAS, that we refer to as synthesizers. Participating agents are assigned a synthesizer agent upon entry to the MAS. 
Synthesizer agents are agents designed with knowledge of the domain context: goals, actions, conflicting states, norms. They are capable of perceiving all the actions 
of the agents for which they are responsible and the environment state at any given time. 
Synthesizer agents in this model are inspired by the assistant agents in~\cite{campos_robust_2013}, which compute regulations utilising a partial perspective of the MAS, but do so after observing a problem, whereas synthesizer agents here await requests from participating agents. 
Additionally, assistants~\cite{campos_robust_2013} vote on the regulations determining which shall be included, and similarly synthesizer agents must vote to decide whether or not to include all the norms proposed after discussions. 

We note that synthesizer agents in our model can potentially be proactive as well. They can be proactive by examining traces and identifying when conflicting states occur, then proposing norms to avoid them in the future, without waiting on instruction from the agent to do so. Such an approach echoes elements of \cite{morales2013automated,morales_online_2015,Morales:2017:ESS:3091125.3091391,morales_off-line_2018}. Synthesizer agents could also have other functions within a MAS, such as being responsible for the enforcement of sanctions if violation occurs. We have however decided to limit the functions of the synthesizers in this framework to  
participating in the process of synthesising norms for the normative system only upon receipt of requests from agents. This we believe allows us to demonstrate norm synthesis initiated from the bottom-up which is different from existing research on norm synthesis.

\medskip
\noindent \textbf{Components of the framework} The stages of the conceptual model of our framework are: \begin{inparaenum}[(i)]
\item norm creation,  
\item norm propagation, 
\item norm adoption, and
\item norm emergence 
\end{inparaenum} as shown in Figure~\ref{fig:summary} and Figure~\ref{fig:model}. We discuss each of these in the remaining sections of the paper. 

\section{Norm Creation stage}
\label{normCreate}
The creation stage comprises three sub-processes \begin{inparaenum}[(i)]
\item ideation
\item norm synthesis, and 
\item decision, depicted in Figure~\ref{fig:model}, component 1.
\end{inparaenum}
In the ideal scenario, at the end of creation, there is a norm that must be incorporated into the normative system. If not, the request would not have been approved, and thereafter the synthesizers and the initial requesting agent must be informed.

\subsection{Ideation}
The ideation process models the norm creation stage of agents in the emergence perspective, Figure~\ref{fig:model}, component 1(a), by enabling the agent to request changes based on feedback from its interaction with the environment and other agents, similar to how an agent would learn a norm in the emergence perspective. The difference here is that the agent itself is not capable of making changes to the normative system, but can communicate with a synthesizer agent to initiate that change on their behalf. 

A participating agent in a society may determine there is a need for a norm change: a new norm or the revision of an existing norm governing the MAS. 
Norm change requests are typically predicated on the following circumstances:
\begin{inparaenum}[(i)]
\item conflict situation/state arising from compliance with prevailing norms,
\item reasoning that repeatedly determines that violation of current norms is a rational choice,
\item a new norm or norm revision that can potentially bring about a better outcome for the agent or a better state in the MAS; we refer to this as an innovation 
norm, 
\item dissatisfaction with the current norms, e.g. prohibition of actions that can help agents achieve goals more efficiently.
\end{inparaenum}

An agent operating within the society may likely recognise one of the above situations developing long before an external observer can do so. The agent then informs their assigned synthesizer agent
of their perceived need for a norm change. Each request 
must specify the context of the request, the reason for the request and the actual norm proposed, if the agent is capable of synthesising the norm. It may be useful to provide a template to the agents which specifies what needs to be included in the request. 
Individual agents may or may not be able to determine what the new norm can be, since they have limited knowledge of the system. 
We could put more responsibility on the participating agents to be able to perform norm synthesis themselves, but this would require 
us to set minimum capabilities, such as their having high cognitive ability and providing them access to knowledge of activities within the MAS. 
The framework is intended to be accessible for most normative agents so we prefer an approach which does not impose significant requirements.  

\subsection{Norm Synthesis Stage}
The norm synthesis process, as depicted in Figure \ref{fig:model}, component 1(b), enables the encoding of the agents' request, which would have been influenced by their behaviour, into an explicit legal norm that can address the issue within the request.
Synthesizer agents, upon receipt of a proposal, begin the process of synthesising a new norm to address the proposal, but additionally, the synthesizer agents engage in discussion and the establishment of a consensus about the request. 
The sub-processes of the norm synthesis stage are:
proposal, synthesis, discussion/deliberation, and agreement, each of which we now discuss in more detail.  

\medskip
\noindent \textbf{Proposal}
After receiving a proposal, synthesiser agents must parse the request in preparation for synthesising one or more norms. 
Agents utilise a provided template to make proposals and the synthesizer will be equipped to interpret it. 
Synthesizers will normally address requests chronologically and will only handle one request at a time; if multiple requests occur over a short space of time, they are queued.
It is at this point that synthesisers have to decide which requests warrant action, by applying a filtering mechanism. Synthesizers could employ an automated rejection process, which returns a message to participating agents when their norm requests have been deemed to not warrant action. We note however that a filtering service could potentially be made available to participating agents, whereby an intermediate mechanism processes the complaint, provides feedback and then the agent decides whether to make a request of the synthesizer agent.

\medskip
\noindent \textbf{Synthesis} 
The act of synthesising or encoding the norm occurs here. This process utilises the originating agent’s request, information about the environment that it can perceive and the domain context. 
Alternatively, if the participating agent is able to synthesise a norm that will meet the identified need and include it in the request, then this process entails the synthesizer agent determining the validity of the norm. 

During synthesis, the synthesizer agent must reason whether the synthesised norm(s) be capable of addressing the need identified by the agent. For example, for a conflict situation, the norm should ensure that once followed, that conflict situation is no longer observed, or for an innovation 
norm, the norm once adopted ensures that the agents are more efficient in accomplishing their goals, and those of the MAS, or the observation of more acceptable states in the MAS.  

Existing literature employs different mechanisms for synthesising norms. \newline SENSE~\cite{morales_off-line_2018} considers the context of the interacting agents in the time-step before the conflict situation and the actions taken by the agents in that time-step. It then synthesises a norm by prohibiting the action of any one of the participating agents. This is similar to IRON~\cite{morales2013automated,morales_online_2015}. Another
technique, utilised in \cite{tingting-thesis,DBLP:conf/dalt/AthakraviCRVPS12}, is built on inductive logic programming and uses the following inputs: a normative system (which could be an empty set), the observation traces and the normative conditions that must hold in the final states. It utilises the proceeding to revise the normative system producing norms (rules) that are compatible with the supplied trace and the condition. Both~\cite{tingting-thesis,DBLP:conf/dalt/AthakraviCRVPS12} and~\cite{morales_off-line_2018} are offline mechanisms, but can potentially be replicated for online use. 
Finally, the proposed new norm is put forward to the other synthesizer agents for discussion and agreement.
 
\medskip
\noindent \textbf{Discussion/Deliberation}
The synthesising agent informs other synthesizer agents about the new norm and solicits a discussion via a discussion mechanism, which we will call a discussion board, 
that is visible and accessible only to synthesizer agents. 
There is potential to have the agents in the system be able to view this discussion board.
We considered the possibility of providing \textit{read-only} access to participating agents, but this would require them to have the ability to parse and understand the discussions on the board, erecting further technical barriers to access. 

Synthesizers are alerted when there is a new activity on the discussion board 
and they can engage in the discussion when ready. 
The presenting synthesizer agent may be required to defend their proposed norm. An argumentation or negotiation scheme may be appropriate here and can be developed based on existing argumentation frameworks such as \cite{riveret_probabilistic_2012,boella_bdi_2003,bench-capon_norms_2017,atkinson_practical_2007}. 
Synthesiser agents will be required to defend the validity of the proposed norm under scrutiny 
to other synthesizers, not unlike the agents in \cite{atkinson_practical_2007}, who have to defend their preferred action or abort. 

The resolution of internal norm conflicts and modifications, based on new perspectives highlighted by the other synthesizers, can result in the revision of the proposed norm.  
It is the responsibility of the synthesizer that proposes the norm to ensure that after any changes made during this process, the modified norm can still meet the needs of the initial request. 
Therefore, it is possible that the initial proposed norm may be revised
, and the actual outcome is a consequence of modifications made as part of this stage. 


\medskip
\noindent \textbf{Agreement} 
To proceed to the next stage, there needs to be agreement, where, say, a predetermined percentage of synthesisers must agree that the new norm should be introduced into the normative system. Synthesizers can come to agreement using several methods: 
\begin{inparaenum}[(i)]
\item a clear direction forward can be established based on the outcome(s) of the previous discussion phase, 
\item synthesizers can attempt to reach a consensus on the norm proposal for or against it proceeding, or 
\item synthesizers can utilise a voting mechanism (agreed in advance). 
\end{inparaenum} 

If a decision to allow the norm proposal to proceed cannot be reached, the presenting synthesizer must inform the proposing agent that the norm change was rejected. 
The synthesizer can provide reasons for the rejection so that the proposing agent can possibly refine and submit a new request/proposal. 
Once a decision is reached, the agreed norm change is passed on to the decision stage for a final decision on its inclusion in the normative system of the MAS.

\subsection{Decision Stage}
\label{decision}
The decision about whether a norm change will be made is done by a decision making mechanism, which we will refer to as the ``Oracle'' as depicted in Figure~\ref{fig:model}, component~1(c). 
The Oracle is assumed to have access to the entire state of the MAS. The context and domain of the MAS may require that this decision incorporates human input \cite{zanzotto_viewpoint_2019,rahwan_society---loop_2018}. 
Human input might be necessary to preclude the risk of destabilisation of the MAS due to repeated norm synthesis or the synthesis of norms that are potentially detrimental to the purpose of the MAS. 
The domain of the MAS being modelled will determine whether the Oracle's final decision to modify the normative system requires human input, or if an automated decision only can be allowed. For example, in a MAS interacting directly with humans or whose impact can be catastrophic, 
it may be preferred if a human(s) authorises the changes to the normative system. 

A possible solution would be for the mechanism to make a recommendation that needs to 
be accepted by a human oversight process 
before implementation. 
We note that though including a human at this stage can provide human accountability, it raises the challenge of ensuring that the human and the system have the same understanding of the goals of the system. Otherwise we risk a situation where both human and system may need to prove which party is correct. This could necessitate further arbitration mechanisms to avoid stagnation within the MAS.

Though norm conflicts would have been considered in discussions with other synthesizers, there may still be norm conflicts arising from external interaction. This can occur when the MAS may be itself governed by a higher level MAS~\cite{king_automated_2017} 
and the norms of this MAS must not conflict with the norms of the governing MAS. 
These conflicts, if they exist must be resolved, and this can be achieved utilising techniques similar to \cite{king_automated_2017}.
The stability of the normative system within the MAS must also be considered and should be incorporated into the final decision on the modification of the normative system. 
At the end of this process, the decision to include or reject the norm proposed is communicated to the synthesizer agents. 

\section{Norm Propagation}
\label{normProp}
At the end of the norm creation stage, a norm has either been accepted for inclusion into the normative system or been rejected. 
The preceding decision must be communicated to the synthesizer agents by the Oracle. The norm propagation activities are depicted in Figure~\ref{fig:model}, component 2. 
In the case of a modification to the normative system, 
synthesiser agents are tasked with spreading this information to the agents they are responsible for as 
it is necessary that all participating agents in the MAS be aware of any changes to the normative system. 
Synthesizers can choose an appropriate 
norm propagation mechanism to communicate with their assigned participating agents. Broadcasting is one solution, a common knowledge source, as used in \cite{savarimuthu_social_2008} is another, which advises agents when new information is available, or synthesizers can use a distributed sharing mechanism, where they inform some agents, who then inform other agents they are connected to. This latter approach 
could possibly affect 
whether all agents become aware of the modifications to the normative system in a timely manner.  
Alternatively, if the norm change proposal was rejected, the synthesizer needs to communicate this to the agent that made the initial proposal.

\section{Norm Adoption}
\label{normAdopt}
The adoption of the norm requires the agent to internalise and reason about adopting the norm. The norm adoption activities are depicted in Figure~\ref{fig:model}, component 3. Internalising the norm means making an internal representation of the norm, 
while the adoption of the norm is the decision whether to adopt the norm after reasoning about it.
The internalisation of the norm is initiated when agents are made aware of modifications to the normative system. 
As agents are assumed to have an explicit internal representation of norms, see Section \ref{description}, this internalisation of the norm means agents will have to incorporate the norm as part of their beliefs. 
Once internalised, agents can 
reason about the validity and applicability of the norm and decide whether it is useful to adopt it.  
Some agents will do this once and adopt the norm in every situation that it is applicable going forward, as with the normative agents in \cite{criado_normative_2010}. Other agents will reason about it every time a decision to act needs to be made, similar to agents in \cite{dos_santos_neto_developing_2011,bench-capon_norms_2017} and graded normative agents in \cite{criado_normative_2010}.  

\section{Norm Emergence}
\label{normEmer}
The norm emergence activities are depicted in Figure~\ref{fig:model}, component 4.
Once an agent has adopted the norm and complies with it, the percentage of agents doing so may reach the threshold for emergence. At this point it can be said that the norm has emerged within the MAS. If the percentage of agents adopting the norm never reaches the threshold for emergence, it may be useful to see if the opposite prevails. That is, where a large percentage of agents violate the norm. If this is the case then it may be necessary to reevaluate the norm's place within the normative system, which could ultimately lead to its removal.

\section{Conclusions} 
\label{conclude}
In this paper we present a framework for norm emergence in normative MAS premised on utilising the experiences of participating agents to trigger changes in the normative system. 
We describe a distinguished population of synthesizer agents, who accept requests from participating agents and synthesise norms in response to these requests. 
We believe that this is a useful approach for open MAS as there is no need to impose any requirements on the participating agents but can instead provide a set of synthesizer agents that have the requisite capabilities to perform the role. 
The introduction of synthesisers, that can propose changes to the normative system, 
provides participating agents with recourse when they are not satisfied with their experience in the MAS. Instead of leaving the MAS, they can potentially initiate changes within the MAS that will not only improve the experience for themselves, but perhaps for other agents. Participating agents will only need to know who their synthesizer agent is and what information should be provided. 

We posit that the use of special-purpose synthesizer agents that address the needs of participating agents, equips the MAS to modify its normative system at runtime, thereby facilitating decentralised runtime norm synthesis in the MAS. 
Furthermore we posit that the inclusion of discussion and agreement phases 
for synthesizer agents to agree on the inclusion of norms is an important addition to the framework. 
A synthesizer will synthesise norms based on a partial context, and as a result it could synthesise norms that, though capable of resolving the proposed issue, are not useful for the collective MAS. 

The remaining synthesiser agents will be able to assess the proposed norm and determine how it will affect their own view 
of the MAS. The intention is that only norms that are beneficial to a majority or all agents of the MAS shall achieve consensus to proceed to the final stage of verification in the MAS: the \textit{decision stage}. 
The goal is that this additional layer will aid the MAS by  
helping to maintain the stability of the normative system and preventing agents from introducing norms that are detrimental to it. 
It is also the stage where norms can be checked for compliance with governing external MASs, if they exist and/or non-negotiable MAS norms/rules. This stage can also allow for human input depending on the domain of the MAS.

Synthesizers then discuss and decide if the proposed norm should be introduced into the normative system. Finally an Oracle mechanism must approve the inclusion of the norm into the MAS. 
The normative system is modified if the norm is approved, and agents within the MAS need to be informed of changes to the normative MAS. 
Participating agents will then reason about adopting the norm and over time a threshold of agents may choose to adopt the norm. Ultimately we may observe the emergence of the norm in the MAS that was synthesised based on a request by a participating agent. 

%
%

%
%
%
\def\doi#1{\href{https://doi.org/#1}{DOI: \nolinkurl{#1}}}
\bibliographystyle{splncs04}
\bibliography{biblio,biblio2}

\end{document}